\documentclass{raa}
\usepackage{txfonts}
\usepackage{graphicx,times}
\usepackage{natbib}
\usepackage{amssymb}
\usepackage{threeparttable}
\begin{document}

   \title{Precise determination of fundamental parameters of six exoplanet host stars and their planets}

   \volnopage{Vol.0 (200x) No.0, 000--000}      
   \setcounter{page}{1}          

   \author{K. Liu
      \inst{1}
   \and S. L. Bi
      \inst{1}
   \and T. D. Li
      \inst{1,2}
   \and Z. E. Liu
      \inst{1}
   \and Z. J. Tian
      \inst{1}
   \and Z. S. Ge
      \inst{1}
   }

   \institute{Department of Astronomy, Beijing Normal University,
             Beijing 100875, China; {\it liukang@mail.bnu.edu.cn; bisl@bnu.edu.cn}\\
        \and
             National Astronomical Observatories, Chinese Academy of Sciences,
             Beijing 100012, China; \\
   }

   \date{Received~~2009 month day; accepted~~2009~~month day}

\abstract{The aim of this paper is to determinate the fundamental
parameters of six exoplanet host (EH) stars and their planets. While
techniques for detecting exoplanets yield properties of the planet
only as a function of the properties of the host star, hence, we
must accurately determine parameters of EH stars at first. For this
reason, we constructed a grid of stellar models including diffusion
and rotation-induced extra-mixing with given ranges of input
parameters (i.e. mass, metallicity, and initial rotation rate). In
addition to the commonly used observational constraints such as the
effective temperature $T_{\rm{eff}}$, luminosity $L$ and metallicity
[Fe/H], we added two observational constraints, the lithium
abundance $\log$ $N$ (Li) and the rotational period $P_{\rm{rot}}$.
These two additional observed parameters can make further constrains
on the model due to their correlations with mass, age and other
stellar properties. Hence, our estimations of fundamental parameters
for these EH stars and their planets are with higher precision than
previous works. Therefore, the combination of rotational period and
lithium help us to obtain more accurate parameters for stars,
leading to an improvement of the knowledge of the physical state
about the EH stars and their planets.  }

   \authorrunning{Precise determination of fundamental parameters of six EH stars and their planets}
   \titlerunning{K. Liu et al.}

   \maketitle

\section{Introduction}
\label{sect:intro}

In the past two decades, thousands of identifiable planets outside
the solar system have already been spotted
\citep[e.g.][]{Vogt00,Pie10,Mou11,Ofir13,Rowe14}. An overwhelming
majority of them have been discovered using indirect methods, i.e.,
radial-velocity (RV) and photometric transits. This is due to the
fact that planets are non-luminous bodies, they merely reflects
light of the parent star. A planet is like a small "undetectable"
speckle in the stellar image, from a distance of a few parsec. But a
planet can cause dynamical perturbations onto parent star, providing
the possibility to detect it by indirect means. As a consequence,
this makes the characteristic parameters of a exoplanet depend
strongly and only on the characteristic parameters of the host star.
Hence, it is of great significance to obtain the accurate masses and
radii of host stars to study exoplanets
\citep[e.g.][]{Sea03,San08,Winn10}.

The most commonly used method for determination of the stellar
properties is to fit the parameters of theoretical models with the
observational constraints, e.g. effective temperature $T_{\rm{eff}}$
and luminosity $L/L_{\sun}$. However, this method generally makes it
difficult to obtain sets of complete stellar atmospheric parameters.
There are a few of free parameters in stellar structure and
evolution models, which bring more uncertainties in the
observations, result in insufficient estimation of the stellar
properties \citep{Bi08}.

Nevertheless, the abundance of lithium in stellar photospheres is
usually used to study various processes, from big bang
nucleosynthesis to the formation and evolution of planetary systems
\citep[e.g.,][]{Mel10,San10}. Lithium is readily destroyed in
stellar interiors at a comparatively low temperature ($\sim$ 2.5
$\times$ 10$^6$ K), therefore in solar-like stars, the surface
lithium abundance was treated as an extremely sensitive diagnostic
for stellar structure and evolution. Additionally, the evolution of
lithium abundance during pre-MS and MS phases have strong
correlations on several properties of star, i.e., metallicity, mass,
age, and rotational history. Meanwhile, due to total angular
momentum loss induced by magnetic braking, the rotational period of
solar-like stars decreases with age during the main sequence. And
the rate of angular momentum loss is related to stellar mass, radii
and rotational rates. Accordingly, by combining non-standard stellar
model which includes the process of extra-mixing with accurate
measurement of lithium abundance and rotational period, we can
obtain more precise estimation about the fundamental parameters of
EH stars and their planets \citep{Do Nascimento09,Cas11,Li12}.

In this work, we computed evolutionary models including
rotation-induced element mixing and microscopic diffusion. We used
three commonly used observation constraints ($T_{\rm{eff}}$,
$L/L_{\sun}$ and [Fe/H]) and two additional observation constraints
(the lithium abundance $\log$ $N$ (Li) and the rotational period
$P_{\rm{rot}}$) as restrictions of stellar models, aiming to provide
the stellar parameters accurately. We assumed that the depletion of
lithium happens at pre-main-sequence (pre-MS) and evolve differently
with different initial rotational rates.

The observed data of the six EH stars and their planets are
summarized in Section ~\ref{sect:Obs}. The computational method and
the details of our evolutionary models are described in Section
~\ref{sect:Mod}. In Section ~\ref{sect:Res}, we present our modeling
results and compare with previous studies. Finally, the discussion

\section{The selection of star sample}
\label{sect:Obs}

\subsection{EH Stars}

\begin{table}
\begin{center}
\caption{Main Characteristics of Six EH Stars.}\label{tbl1}
\begin{tabular}{cccccccc}
  \hline\noalign{\smallskip}
HIP & HD & $T_{\rm{eff}}$ & $\log(L/L_{\sun})$ & [Fe/H] & $\log$ $N$ (Li) & $P_{\rm{rot}}$& ref \\
    &    & (K)            &                    &        & (dex)           &               &     \\
  \hline\noalign{\smallskip}
    9683   & 12661  &  5743$\pm$44 &  0.093$\pm$0.063  &   0.36$\pm$0.03  & ...           & ...        & (1)  \\
           &        &  5785$\pm$50 &  0.033$\pm$0.063  &   0.37$\pm$0.03  & 1.10$\pm$0.60 & ...        & (2)  \\
           &        &  5715$\pm$70 &  ...              &   0.36           & ...           & ...        & (3)  \\
           &        &  ...         &  ...              &   ...            & ...           & 35$\pm$2.1 & (5)  \\
           &        &  5748$\pm$70 &  0.063$\pm$0.063  &   0.36$\pm$0.03  & 1.10$\pm$0.60 & 35$\pm$2.1 & (6)  \\
    \ & & & & & & & \\
    33212  & 50554  &  5929$\pm$44 &  0.167$\pm$0.063  &  -0.07$\pm$0.03  & ...           & ...        & (1)  \\
           &        &  5982$\pm$26 &  0.119$\pm$0.062  &  -0.07$\pm$0.02  & 2.40$\pm$.011 & ...        & (2)  \\
           &        &  6050$\pm$70 &  ...              &   0.02           & 2.59          & ...        & (3)  \\
           &        &  ...         &  ...              &   ...            & ...           & 16$\pm$1.0 & (5)  \\
           &        &  5987$\pm$70 &  0.143$\pm$0.063  &  -0.04$\pm$0.03  & 2.50$\pm$0.11 & 16$\pm$1.0 & (6)  \\
    \ & & & & & & & \\
    47007  & 82943  &  5997$\pm$44 &  0.169$\pm$0.047  &   0.27$\pm$0.03  & ...           & ...        & (1)  \\
           &        &  6011$\pm$36 &  0.152$\pm$0.061  &   0.28$\pm$0.03  & 2.47$\pm$0.10 & ...        & (2)  \\
           &        &  6025$\pm$70 &  ...              &   0.33           & 2.52          & ...        & (3)  \\
           &        &  ...         &  ...              &   ...            & ...           & 20$\pm$1.2 & (5)  \\
           &        &  6011$\pm$70 &  0.161$\pm$0.061  &   0.29$\pm$0.03  & 2.50$\pm$0.10 & 20$\pm$1.2 & (6)  \\
    \ & & & & & & & \\
    50473  & 89307  &  5898$\pm$44 &  0.096$\pm$0.062  &  -0.16$\pm$0.03  & ...           & ...        & (1)  \\
           &        &  5914$\pm$25 &  0.130$\pm$0.063  &  -0.18$\pm$0.02  & 2.18$\pm$0.11 & ...        & (2)  \\
           &        &  ...         &  ...              &   ...            & ...           & 18$\pm$1.1 & (5)  \\
           &        &  5906$\pm$44 &  0.113$\pm$0.063  &  -0.17$\pm$0.03  & 2.18$\pm$0.11 & 18$\pm$1.1 & (6)  \\
    \ & & & & & & & \\
    59610  & 106252 &  5870$\pm$44 &  0.107$\pm$0.069  &  -0.08$\pm$0.03  & ...           & ...        & (1)  \\
           &        &  5923$\pm$38 &  0.108$\pm$0.063  &  -0.05$\pm$0.03  & 1.69$\pm$0.13 & ...        & (2)  \\
           &        &  5890$\pm$70 &  ...              &  -0.01           & 1.65          & ...        & (3)  \\
           &        &  5899$\pm$62 &  ...              & -0.034$\pm$0.041 & 1.71$\pm$0.04 & ...        & (4)  \\
           &        &  ...         &  ...              &   ...            & ...           & 23$\pm$1.4 & (5)  \\
           &        &  5896$\pm$70 &  0.108$\pm$0.069  &  -0.04$\pm$0.04  & 1.68$\pm$0.13 & 23$\pm$1.4 & (6)  \\
    \ & & & & & & & \\
    77740  & 141937 &  5847$\pm$44 &  0.070$\pm$0.073  &   0.13$\pm$0.03  & ...           & ...        & (1)  \\
           &        &  5842$\pm$36 & -0.026$\pm$0.067  &   0.10$\pm$0.03  & 2.26$\pm$0.11 & ...        & (2)  \\
           &        &  5925$\pm$70 & ...               &   0.11           & 2.48          & ...        & (3)  \\
           &        &  5900$\pm$19 & ...               &  0.125$\pm$0.030 & 2.36$\pm$0.02 & ...        & (4)  \\
           &        &  ...         &  ...              &   ...            & ...           & 21$\pm$1.3 & (5)  \\
           &        &  5879$\pm$70 &  0.022$\pm$0.073  &   0.12$\pm$0.03  & 2.37$\pm$0.11 & 21$\pm$1.3 & (6)  \\
  \noalign{\smallskip}\hline
\end{tabular}
\begin{list}{}{}
\item[$^{\mathrm{}}$] Reference: (1) \cite{VF05}; (2) \cite{Ghe10a} and \cite{Ghe10b}; (3)
\cite{Isr04}; (4)\cite{Bau10}; (5)\cite{Wri04}; (6) Mean value.
\end{list}
\end{center}
\end{table}

The sample stars for our modeling are six solar-analog stars with
observed lithium abundances and rotational periods, and their
planets have been detected by using radial-velocity. The detections
of lithium abundances and rotational periods make it is possible to
obtain precise estimations of stellar parameters, especially the
masses and radii which are significant for us to determine the
properties of their planets.

We summarized the observed data of EH stars which were used for our
theoretical calculations in Table ~\ref{tbl1}. The atmospheric
features, $T_{\rm{eff}}$, $L$, and [Fe/H] were collected from
\citet{VF05}, \citet{Ghe10a}, \citet{Isr04} and \citet{Bau10}. We
adopted the lithium abundance $\log$ $N$(Li) from the observations
of \citet{Ghe10b}, \citet{Isr04} and \citet{Bau10}. The rotational
period $P_{\rm{rot}}$ we used was determined by \citet{Wri04} from
the California and Carnegie Planet Search Program with the HIRES
spectrometer at Keck Observatory. The average values of these
observations were adopted in the following study.

The Spectra of \citet{VF05} were obtained with the HIRES
spectrograph mounted on the 10-m telescope at Keck Observatory
\citep{Vogt94}, the UCLES spectrograph mounted on the 4-m
Anglo-Australian Telescope at Siding Spring Observatory
\citep{Die90}, and the Hamilton echelle spectrometer at Lick
Observatory \citep{Vogt87}. The spectra of \citet{Ghe10a,Ghe10b}
were obtained with the FEROS spectrograph mounted on the MPG/ESO
2.20-m telescope at La Silla \citep{Kau99}. The observations of
\citet{Isr04} were carried out using the UES/4.2-m William Hershel,
the SARG/3.5-m TNG at La Palma, and the FEROS/1.52-m ESO, the
CORALIE/1.2-m Euler Swiss at La Silla. Stars from \citet{Bau10} were
observed with RGT spectrograph mounted on the 2.7-m Harlan Smith
telescope at the McDonald observatory, MIKE spectrograph mounted on
the 6.5-m Magellan Clay telescope at Las Campanas observatory, and
HARPS spectrograph mounted on the 3.6-m ESO telescope at La Silla
observatory.

\subsection{Exoplanets}

\begin{table}
\begin{center}
\caption{Main Characteristics of Exoplanets.}\label{tbl5}
\begin{tabular}{lcccc}
  \hline\noalign{\smallskip}
    Planet & $P$ & $e$ & $K_1$ & Ref \\
           &  (day)           &              &   (m s$^{-1}$)     &     \\
  \hline\noalign{\smallskip}
   HD 12661b  &  262.709 $\pm$ 0.083 &  0.3768 $\pm$ 0.0077  &   73.56 $\pm$ 0.56  &   (1)  \\
   HD 12661c  &  1708.0 $\pm$ 14.0   &  0.031 $\pm$ 0.022    &   30.41 $\pm$ 0.62  &   (1)  \\
   HD 50554b  &  1293.0 $\pm$ 37.0   &  0.501 $\pm$ 0.030    &   104 $\pm$ 5       &   (2)  \\
   HD 82943b  &  442.4 $\pm$ 3.1     &  0.203 $\pm$ 0.052    &   39.8 $\pm$ 1.3    &   (3)  \\
   HD 82943c  &  219.3 $\pm$ 0.8     &  0.425 $\pm$ 0.018    &   54.4 $\pm$ 2.0    &   (3)  \\
   HD 82943d  &  1072 $\pm$ 13       &  0 $\pm$ 0            &   5.39 $\pm$ 0.57   &   (4)  \\
   HD 89307b  &  2199 $\pm$ 61       &  0.25 $\pm$ 0.09      &   32.4 $\pm$ 4.5    &   (5)  \\
   HD 106252b &  1600.0 $\pm$ 18.0   &  0.471 $\pm$ 0.028    &   147 $\pm$ 4       &   (2)  \\
   HD 141937b &  653.22 $\pm$ 1.21   &  0.41 $\pm$ 0.01      &   234.5 $\pm$ 6.4   &   (6)  \\
  \noalign{\smallskip}\hline
\end{tabular}
\begin{list}{}{}
\item[$^{\mathrm{}}$] Reference: (1) \cite{Wri09}; (2) \cite{Per03}; (3) \cite{Tan13}; (4) \cite{Bal14}; (5) \cite{Boi12}; (6) \cite{Udr02}.
\end{list}
\end{center}
\end{table}

We listed the planetary orbital parameters which were obtained by RV
measurements in Table~\ref{tbl5}. Two of these systems are found
multi-planetary. HD 12661 and HD 82943 host two and three planets,
respectively. Orbital period $P$, eccentricity $e$, and
semi-amplitude $K_1$ (the velocity wobble) were gave here, for more
details about the planetary such as the periastron passage time $T$
and the angle between the periastron and the line-of-nodes $\omega$
can be found in the related literatures.

The radial velocity data were from the HIRES spectrograph (HD 12661
and HD 82943) mounted on 10-m Keck-1 telescope at the Keck
Observatory \citep{Vogt94}, the ELODIE echelle spectrograph (HD
50554 and HD 106252) and the SOPHIE spectrograph (HD 89307) mounted
on the Cassegrain focus of the 1.93-m telescope at Haute-Provence
Observatory \citep{Bar96}, and the CORALIE echelle spectrograph (HD
141937) mounted on the 1.2-m Euler Swiss telescope at La Silla
Observatory \citep{Que00,Udr00}, respectively.

\section{Stellar Models}
\label{sect:Mod}
\subsection{Input Physics}
To estimate the parameters of the sample stars, a grid calculation
were carried out based on a stellar evolutionary model named Yale
Rotating Stellar Evolution Code (YREC) \citep{Pin90,Pin92,Demarque},
which includes diffusion, angular momentum loss, angular momentum
transport and rotation-induced elements mixing. Detailed
descriptions of model can be found in \cite{Guen92}, \cite{Cha95}
and \cite{Li03}. The calculations were carried out with the
up-to-date OPAL equation-of-state tables EOS2005 \citep{Rogers}.
Solar mixture of GS98 \citep{Grevesse} ($Z_{\sun}$ = 0.0170 and
$(Z/X)_{\sun}$ = 0.0230) was adopted and hence the opacities were
generated with the composition of GS98 \citep{Grevesse} and
supplemented by low-temperature opacities from \citet{Ferguson}.
Atmosphere of the model follows the Eddington $T-\tau$ relation. We
used NACRE reaction rates \citep{Angulo} for nuclear reaction and
the mixing length theory \citep{Bohm} for convection. Following the
formulation of \citet{Thoul}, the gravitational settling of helium
and heavy elements is considered in the stellar model.

When rotation is taken into account, the characteristics of a model
depend on six parameters: mass $M$, age $t$, mixing-length parameter
$\alpha \equiv l/Hp$, two parameters: $X_{\rm{ini}}$ and
$Z_{\rm{ini}}$ described the initial chemical composition of a star,
and rotational period $P_{\rm{rot}}$. To reproduce the evolution of
lithium, evolutions of stars during Pre-MS stage is considered and
hence we selected the initial model for each calculation on the
Hayashi Line. All of the models evolved to exhaust its supply of the
hydrogen in the core. Initial helium abundance ($Y_{\rm{ini}}$ =
0.275) and the mixing-length parameter ($\alpha$ = 1.75) were
regarded as constants in the grid computation.

The ranges of variable parameters of the grid calculation and their
step sizes are shown in Table ~\ref{tbl2}. According to the
effective temperatures of the sample stars, we set the range of mass
from 0.90 to 1.10 $M_{\sun}$ with a grid size of 0.01 $M_{\sun}$.
The range of mass fraction of all heavy elements $Z_{\rm{ini}}$,
which was derived from $Z_{\sun}$ and the observed [Fe/H], is from
0.010-0.040 dex with a grid size of 0.001 dex. Although the initial
models were selected on the Hayashi Line, we used the rotational
rates at zero age main sequence ($V_{\rm{ZAMS}}$) to represent the
rotational conditions for better understanding. The range of
$V_{\rm{ZAMS}}$ is from 20 to 70 km s$^{-1}$, in steps of 10 km
s$^{-1}$.

\begin{table}
\begin{center}
\caption{Input Parameters for Theoretical Calculation.}\label{tbl2}
\begin{tabular}{cccc}
    \hline\noalign{\smallskip}
Variate                            & MIN & MAX & $\delta^{\rm{a}}$ \\
    \hline\noalign{\smallskip}
$M (M_{\sun})$                     & 0.90    & 1.10    & 0.01              \\
Z                                  & 0.010   & 0.040   & 0.001             \\
$V_{\rm{ZAMS}}$ (km $\rm{s}^{-1})$ & 20      & 70      & 10                \\
    \noalign{\smallskip}\hline
\end{tabular}
\end{center}
\tablecomments{0.56\textwidth}{$^{\rm{a}}$ The value of $\delta$
represents the increment between the minimum and maximum values.}
\end{table}

\subsection{Angular Momentum Loss}
The braking law of \citet{Kaw88} is adopted as the angular momentum
loss equation:
\begin{equation}
\frac{{dJ}}{{dt}} = \left\{ \begin{array}{l}
- K{\Omega ^3}{({R}/{{{R_ {\sun} }}})^{1/2}}{({M}/{{{M_ {\sun} }}})^{ - 1/2}}  (\Omega  \le {\Omega _{sat}}) \\
\\
- K\Omega {\Omega _{sat}}^2{({R}/{{{R_ {\sun} }}})^{1/2}}{({M}/{{{M_ {\sun} }}})^{ - 1/2}}   (\Omega  > {\Omega _{sat}}), \\
\end{array} \right.
\end{equation}
where the parameter $K$ is constant for all stars, and associated
with the magnetic field intensity. $\Omega_{\rm{sat}}$ is the
angular velocity of surface when the magnetic saturation occurs in
star. Both $K$ and $\Omega_{\rm{sat}}$ are free parameters and we
follow \citet{Bou97} setting $K$ = 2.0 $\times$ $10^{47}$$\rm{g}$
$\rm{cm^2}$ $\rm{s}$ and $\Omega_{\rm{sat}}$ = 14 $\Omega_{\sun}$.

\subsection{Extra-mixing in the Radiative Region}
Besides the microscopic diffusion of of elements, which we have
mentioned above, angular momentum transport and elements mixing
caused by rotation are taken into account in radiative regions.
These processes can be described as a couple of diffusion equations
\citep{Cha95}:
\begin{equation}
\rho {r^2}\frac{I}{M}\frac{{d\Omega }}{{dt}} = \frac{d}{{dr}}(\rho
{r^2}\frac{I}{M}{D_{rot}}\frac{{d\Omega }}{{dt}}),
\end{equation}
\begin{equation}
\rho {r^2}\frac{{d{X_i}}}{{dt}} = \frac{d}{{dr}}\left[\rho
{r^2}{D_{m,1}}{X_i} + \rho {r^2}({D_{m,2}} +
{f_c}{D_{rot}})\frac{{d{X_i}}}{{dt}}\right],
\end{equation}
where $\Omega$ is the angular velocity, $X_i$ is the mass fraction
of chemical species $i$, and $I/M$ is the moment of inertia per unit
mass. $D_{m,1}$ and $D_{m,2}$ are the microscopic diffusion
coefficients. $D_{rot}$ is the diffusion coefficient caused by
rotation-induced mixing. More details of these diffusion
coefficients were given by \citet{Cha95}. The tunable parameter
$f_c$ was used to alter the effects of rotation-induced element
mixing. It was determined by observations, that is, the depletion of
lithium in our solar model must fits the observed depletion in Sun
\citep{Cha95}.

\section{Results}
\label{sect:Res}

\subsection{Stellar Parameters}

\begin{figure}
\centering
\includegraphics[width=\textwidth, angle=0]{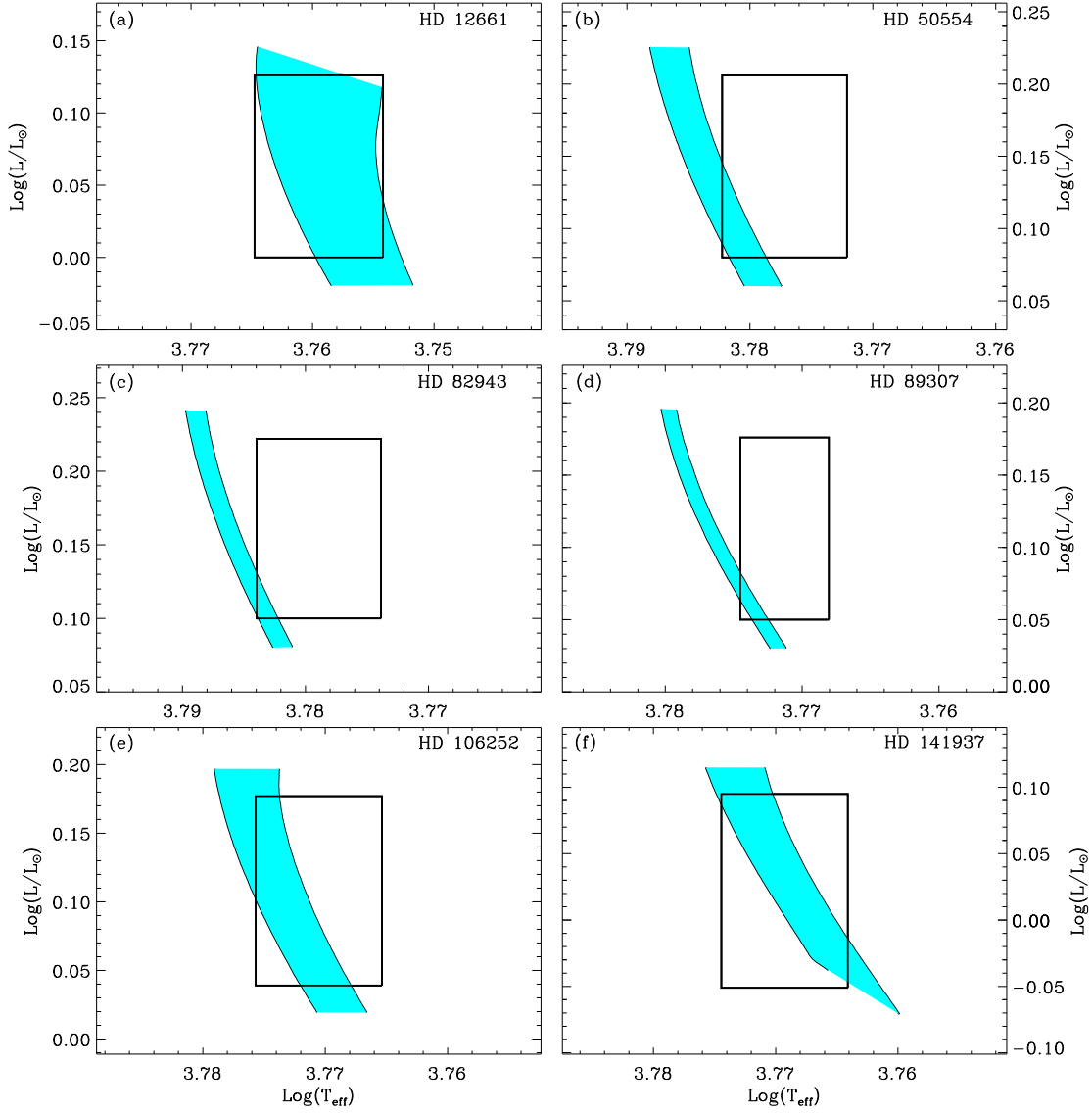}
\caption{Evolutionary tracks of HD 12661, HD 50554, HD 82943, HD
89307, HD 106252, and HD 141937 in the H-R diagram constrained by a)
$T_{\rm{eff}}$ + $L$ + [Fe/H]; b) $T_{\rm{eff}}$ + $L$ + [Fe/H] +
log N(Li); c) $T_{\rm{eff}}$ + $L$ + [Fe/H] + $\log$ N(Li) +
$P_{\rm{rot}}$.} \label{Fig:1}
\end{figure}

we calculated a series of evolutionary models in estimated $M$ and
$Z_{ini}$ ranges to reproduce the observational constraints of these
six EH stars. As shown in Fig. ~\ref{Fig:1}, evolutionary tracks for
each stars are plotted in conformity to observational constraints.
For the sake of simplicity, the situations of star HD 12661 is taken
as an example.

First of all, three classical observed features, the effective
temperature $T_{\rm{eff}}$, luminosity $L$ and metallicity [Fe/H],
were considered and 157 tracks are found fitting these three
observational constraints. The mass and age of HD 12661 provided by
the models are 1.02 $\pm$ 0.03 $M_{\sun}$ and 6.76 $\pm$ 4.31 Gyr.

Secondly, lithium abundance was taken into account. Lithium is the
most important element since it is readily burned in the stellar
interiors. The abundance of lithium indicates the extent of element
mixing in the stars, in addition, the depletion of lithium depends
strongly on the mass and age of star \citep{Do Nascimento09,Li12}.
In this step, there are only 76 evolutionary tracks which fit four
observational constraints, including lithium abundance $\log$ $N$
(Li), and we estimate the mass and age of the star HD 12661 are 1.02
$\pm$ 0.02 $M_{\sun}$ and 5.56 $\pm$ 3.01 Gyr. Additionally, lithium
abundance narrows the ranges of input parameters, thus the possible
position of the star in the H-R diagram is restricted to a smaller
field than what has been obtained above.

Finally, after adding the rotational period to our models as a
constraint, only 30 evolutionary tracks are found fitting the
observed $P_{\rm{rot}}$. The range of $V_{\rm{ZAMS}}$ is
significantly reduced to 30 km s$^{-1}$ - 40 km s$^{-1}$. In a same
way as lithium abundance, the ranges of input parameters of the
stellar models are also reduced by the rotational period, and hence
it makes further constraining to the possible position of the star
in the H-R diagram as shown in Fig.~\ref{Fig:1}a. The rotational
period $P_{\rm{rot}}$ helps us determinate the mass and age of HD
12661 even more precisely, which are 1.02 $\pm$ 0.02 $M_{\sun}$ and
6.39 $\pm$ 1.94 Gyr.

The same method were adopted for all the other EH stars, we plotted
their evolutionary tracks in Fig.~\ref{Fig:1}, each line illustrate
the each star. Comparing the situation of star HD 12661 with others,
we find that this star occupy larger area in H-R diagram than five
other stars in Fig.~\ref{Fig:1}, this is owing to a large error in
the abundance of lithium for HD 12661.

\subsection{Comparison with Previous Results}

\begin{figure}
\centering
\includegraphics[width=\textwidth, angle=0]{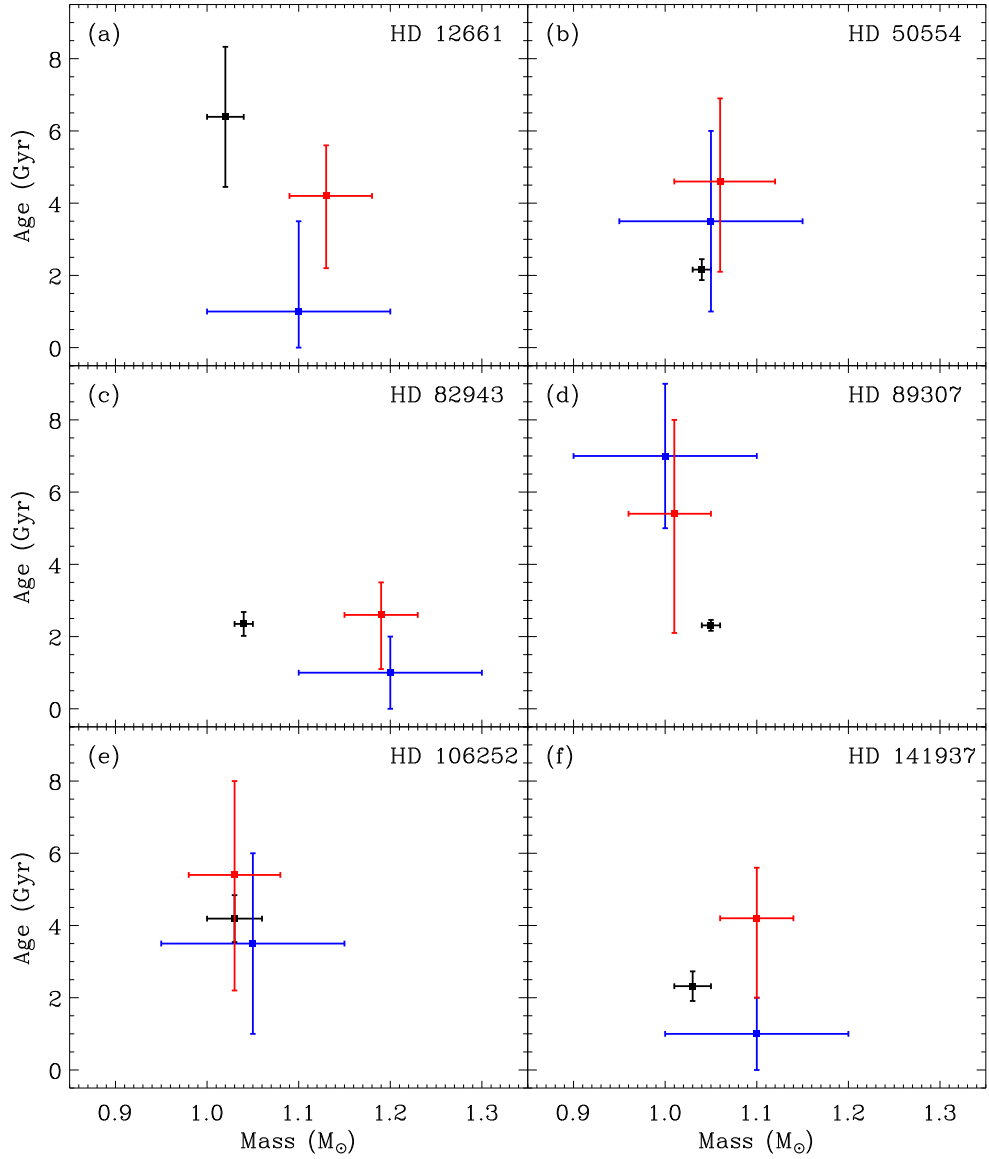}
\caption{Comparisons between masses and ages determined by our model
(the black error bar) and estimates of previous studies for all six
EH stars. The red and blue error bars represent the results of
\cite{VF05} and \cite{Ghe10a}, respectively.}\label{Fig:3}
\end{figure}

These six EH stars were previously studied by several researchers,
the methods and estimates of two of them could be seen in
Table~\ref{tbl4}. \citet{Ghe10a} and \citet{VF05} observed these
stars and provided their masses, radii and ages through different
methods. We compare their results obtained by interpolating
isochrones with ours in the following paragraphs. The comparisons of
masses and ages of these six EH stars were plotted in Fig.
~\ref{Fig:3}.

For the six EH stars, the results of \citet{Ghe10a} given the error
of mass is $\sim$ 0.10 M$_{\sun}$ and the error of age is $\sim$ 2.0
Gyr. The mass determinations of \citet{VF05} were close to those of
\citet{Ghe10a} but with higher precision, i.e., $\triangle M$ $\sim$
0.05 M$_{\sun}$. Our mass estimations of HD 12661, HD 50554, HD
82943, HD 89307, HD 106252, and HD 141937 are $1.02 \pm 0.02
M_{\sun}$, $1.04 \pm 0.01 M_{\sun}$, $1.04 \pm 0.01 M_{\sun}$, $1.05
\pm 0.01 M_{\sun}$, $1.03 \pm 0.03 M_{\sun}$, and $1.03 \pm 0.02
M_{\sun}$, respectively, most of which are less massive than what
have been obtained by \citet{Ghe10a} and \citet{VF05}. This result
is due to the element transport which caused by the interaction
between diffusion and rotation-induced mixing in the stellar
radiative region \citep{Cha95,Egg10}. The process of element
transport changes the chemical composition of the external layers
and hence cause the evolutionary tracks shift to hot side on the H-R
diagram. Thus, when the observed effective temperature is given,
rotational model tends to provide less massive result of mass than
those obtained by standard model. The precision of our mass
determinations is the best of the three, which is 0.01 $\sim$ 0.03.

Ages of these six EH stars provided by \citet{VF05} are mostly older
than those of \citet{Ghe10a}, with a similar accuracy, i.e.,
$\triangle t$ $\sim$ 2.0 Gyr. Our age determinations generally agree
within the errors of previous works, and which are much more
accurate ($\triangle t$ $\sim$ 0.5 Gyr) than determined by
interpolating isochrones. (see Table~\ref{tbl4}). Moreover,
Combining the rotational periods listed in Table~\ref{tbl1} with
ages obtained by us, we found that there is a positive correlation
between them.

This result is reasonable, because the depletion of lithium is a
function of stellar mass, age, rotational rates and metallicity,
while the rotational period increases with age during the main
sequence. Therefore, these two additional observational constraints
can effectively restrict the ranges of input parameter and improve
the precision of the stellar model.

\begin{table}
 \centering
  \caption{Stellar Parameters and Comparison with Previous Studies.}\label{tbl4}
\begin{tabular}{cccccc}
\hline\noalign{\smallskip}
Star & $M$          &  t       &  $\emph{R}$  &Method &Ref.\\
     & ($M_{\sun}$) & (Gyr)    & ($R_{\sun}$) &       &    \\
 \hline\noalign{\smallskip}
 HD 12661   &0.96$\pm$0.47            &...                  &1.04$\pm$0.08   & Spectroscopic          & (1)\\
            &1.10$\pm$0.10            &$1.0^{+2.5}_{-1.0}$  &...             & Isochrones             & (1)\\
            &1.22$\pm$0.18            &...                  &1.124$\pm$0.037 & Spectroscopic          & (2)\\
            &$1.13^{+0.05}_{-0.04}$   &$4.2^{+1.4}_{-2.0}$  &...             & Isochrones             & (2)\\
            &1.02$\pm$0.02            &6.39$\pm$1.94        &1.11$\pm$0.08   & This work              &    \\
       &    &    &    &    &    \\
 HD 50554   &0.81$\pm$0.39            &...                  &1.07$\pm$0.08   & Spectroscopic          & (1)\\
            &1.05$\pm$0.10            &$3.5^{+2.5}_{-2.5}$  &...             & Isochrones             & (1)\\
            &0.93$\pm$0.14            &...                  &1.149$\pm$0.039 & Spectroscopic          & (2)\\
            &$1.06^{+0.06}_{-0.05}$   &$4.6^{+2.3}_{-2.5}$  &...             & Isochrones             & (2)\\
            &1.04$\pm$0.01            &2.16$\pm$0.29        &1.02$\pm$0.02   & This work              &    \\
       &    &    &    &    &    \\
 HD 82943   &1.03$\pm$0.50            &...                  &1.10$\pm$0.09   & Spectroscopic          & (1)\\
            &1.20$\pm$0.10            &$1.0^{+1.0}_{-1.0}$  &...             & Isochrones             & (1)\\
            &1.22$\pm$0.17            &...                  &1.125$\pm$0.029 & Spectroscopic          & (2)\\
            &$1.19^{+0.04}_{-0.04}$   &$2.6^{+0.9}_{-1.5}$  &...             & Isochrones             & (2)\\
            &1.04$\pm$0.01            &2.35$\pm$0.33        &1.03$\pm$0.02   & This work              &    \\
       &    &    &    &    &    \\
 HD 89307   &0.85$\pm$0.41            &...                  &1.11$\pm$0.09   & Spectroscopic          & (1)\\
            &1.00$\pm$0.10            &$7.0^{+2.0}_{-2.0}$  &...             & Isochrones             & (1)\\
            &0.91$\pm$0.13            &...                  &1.069$\pm$0.035 & Spectroscopic          & (2)\\
            &$1.01^{+0.04}_{-0.05}$   &$5.4^{+2.6}_{-3.3}$  &...             & Isochrones             & (2)\\
            &1.05$\pm$0.01            &2.31$\pm$0.15        &1.01$\pm$0.01   & This work              &    \\
       &    &    &    &    &    \\
HD 106252   &1.16$\pm$0.57            &...                  &1.08$\pm$0.09   & Spectroscopic          & (1)\\
            &1.05$\pm$0.10            &$3.5^{+2.5}_{-2.5}$  &...             & Isochrones             & (1)\\
            &1.01$\pm$0.15            &...                  &1.093$\pm$0.040 & Spectroscopic          & (2)\\
            &$1.03^{+0.05}_{-0.05}$   &$5.4^{+2.6}_{-3.2}$  &...             & Isochrones             & (2)\\
            &1.03$\pm$0.03            &4.19$\pm$0.65        &1.05$\pm$0.01   & This work              &    \\
       &    &    &    &    &    \\
HD 141937   &0.58$\pm$0.28            &...                  &0.95$\pm$0.08   & Spectroscopic          & (1)\\
            &1.10$\pm$0.10            &$1.0^{+1.0}_{-1.0}$  &...             & Isochrones             & (1)\\
            &1.07$\pm$0.13            &...                  &1.056$\pm$0.039 & Spectroscopic          & (2)\\
            &$1.10^{+0.04}_{-0.04}$   &$4.2^{+1.4}_{-2.2}$  &...             & Isochrones             & (2)\\
            &1.03$\pm$0.02            &2.32$\pm$0.41        &0.99$\pm$0.06   & This work              &    \\
       &    &    &    &    &    \\
 \noalign{\smallskip}\hline
\end{tabular}
\begin{list}{}{}
\item[$^{\mathrm{}}$] Reference: (1) \cite{Ghe10a}; (2) \cite{VF05}.
\end{list}
\end{table}

\subsection{Planetary Parameters}
For a given planetary system with known orbital parameters, we can
calculate the mass function\citep{San08}:
\begin{equation}\label{eq1}
f(m) = \frac{(M_2\sin i)^3}{(M_1+M_2)^2} = 1.036 \times
10^{-7}K_1^3(1-e^2)^{(3/2)}P
\end{equation}
where $M_1$ and $M_2$ are the masses of the star and planet, $i$ is
the inclination of the line of sight with respect to the orbital
axis, $K_1$ is the semi-amplitude of radial-velocity of the star
with mass $M_1$, $e$ is the orbital eccentricity, and $P$ is orbital
period.

Furthermore, as we know, from Kepler's third law:
\begin{equation}\label{eq2}
\frac{a^3}{P^2} = \frac{G(M_1+M_2)}{4\pi^2}
\end{equation}
where \textit{a} is orbital semimajor axis and \textit{G} is the
universal gravitational constant.

From Equation~\ref{eq1} and ~\ref{eq2}, and combined stellar masses
determined in the previous section, the minimum masses $M_2\sin i$
and orbital semimajor axes $a$ of planets can be obtained, as shown
in Table~\ref{tbl6}. It should be noted that the uncertainty of our
estimation consists of two parts. One is associated with the
observation, such as the errors of $P$, $e$, and $K_1$ (listed in
Table~\ref{tbl5}). The other is produced by the model, specifically,
the error of stellar mass. We summarized the two parts of the
uncertainty separately in Table~\ref{tbl6}. Compared with the
results of previous studies, our determinations are more accurate,
whether including the errors of observations or not.

\cite{Bat13} pointed out that a 0.1 $M_{\sun}$ companion would
induce a systematic error of approximately 2\%. Correspondingly, the
accuracy and precision of parameters of EH star will have a huge
impact on our estimates of properties of planet. Therefore, the
accurate knowledge of EH star is extremely important for the study
of exoplanets.

\begin{table}
\begin{center}
\caption{Planetary Parameters and Comparison with Previous
Studies.}\label{tbl6}
\begin{tabular}{l|ccc|cccccc}
  \hline\noalign{\smallskip}
           & \multicolumn{3}{c|}{Previous Studies} & \multicolumn{6}{c}{This Work}\\
    Planet &  $M\sin i$  & $a$ & Ref &  $M\sin i$   & $\delta_M^{obs}$ & $\delta_M^{theo}$ & $a$ & $\delta_a^{obs}$ & $\delta_a^{theo}$\\
           & $(M_{\rm{Jup}})$ & (AU)&     & $(M_{\rm{Jup}})$ & $(M_{\rm{Jup}})$ & $(M_{\rm{Jup}})$ & (AU) & (AU) & (AU)  \\
  \hline\noalign{\smallskip}
   HD 12661b  &  2.30 $\pm$ 0.19   &  0.831 $\pm$ 0.048       & (1) & 2.176 & 0.024 & 0.028 & 0.8079 & 0.0001 & 0.0052 \\
   HD 12661c  &  1.92 $\pm$ 0.16   &  2.90  $\pm$ 0.17        & (1) & 1.812 & 0.042 & 0.023 & 2.8145 & 0.0153 & 0.0182 \\
   HD 50554b  &  5.16              &  2.41                    & (2) & 4.954 & 0.388 & 0.031 & 2.3530 & 0.0446 & 0.0075 \\
   HD 82943b  &  1.59              &  1.1866                  & (3) & 1.500 & 0.067 & 0.009 & 1.1510 & 0.0053 & 0.0036 \\
   HD 82943c  &  1.58              &  0.7423                  & (3) & 1.500 & 0.071 & 0.009 & 0.7209 & 0.0017 & 0.0023 \\
   HD 82943d  &  0.294 $\pm$ 0.031 &  2.137 $\pm$ 0.017       & (4) & 0.278 & 0.030 & 0.001 & 2.0766 & 0.0167 & 0.0066 \\
   HD 89307b  &  2.0 $\pm$ 0.4     &  3.34 $\pm$ 0.17         & (5) & 2.074 & 0.356 & 0.013 & 3.3632 & 0.0619 & 0.0106 \\
   HD 106252b &  7.56              &  2.70                    & (2) & 7.613 & 0.364 & 0.147 & 2.7033 & 0.0202 & 0.0259 \\
   HD 141937b &  9.7               &  1.52                    & (6) & 9.316 & 0.306 & 0.120 & 1.4877 & 0.0018 & 0.0095 \\
  \noalign{\smallskip}\hline
\end{tabular}
\begin{list}{}{}
\item[$^{\mathrm{}}$] Reference: (1) \cite{Wri09}; (2) \cite{Per03}; (3) \cite{Tan13}; (4) \cite{Bal14}; (5) \cite{Boi12}; (6) \cite{Udr02}.
\end{list}
\end{center}
\end{table}

\section{Discussion and Conclusion}
\label{sect:Con}

We made an investigation of the physical state of the six EH stars
and their own planet, by employing the method presented by \citet{Do
Nascimento09}. In the context of commonly observations we added two
observational constraints, the lithium abundance $\log$ $N$ (Li) and
the rotational period $P_{\rm{rot}}$, as constraints to better
determine the fundamental parameters of EH stars and their planets.

We gave the estimations of stellar masses and ages using only the
effective temperature $T_{\rm{eff}}$ and luminosity $L/L_{\sun}$ as
observation constraints. The uncertainties of the mass and age are
approximately 0.05 $M_{\sun}$ and 4.0 Gyr. As we considered the
lithium abundance $\log$ $N$ (Li) and rotational period
$P_{\rm{rot}}$ in our analysis, we obtained more precise
determinations. The lithium abundance helped us to minimize the
errors of masses and ages to 0.03 $M_{\sun}$ and 3.0 Gyr,
respectively. Additionally, we used the rotational period
$P_{\rm{rot}}$ to restrict stellar models based on the former
results. The precision has been improved with $\Delta$M $\sim$ 0.02
$M_{\sun}$ and $\Delta$t $\sim$ 0.5 Gyr. Furthermore, because of the
precise determination of age, we restricted atmospheric
characteristics more strictly than the observations, and positioned
the stars more exactly in the H-R diagram. Furthermore, we obtained
the accurate planetary parameters, i.e., minimum masses $M_2\sin i$
and orbital semimajor axes $a$ by using RV measurements and stellar
masses determined previously.

If we want to completely characterize a system, and obtain accurate
properties of the planet, i.e., the mass, radius, and density, we
need the photometric transit, the RV observations and the properties
of EH star. In the future, we hope to conduct further studies with
the data from Gaia mission.

\begin{acknowledgements}
This work is supported by the grants 10933002, 11273007 and 11273012
from the National Natural Science Foundation of China, and the
Fundamental Research Funds for the Central Universities.
\end{acknowledgements}

\label{lastpage}

\end{document}